\documentclass[aps,showpacs,preprintnumbers,showkeys, amsmath, amssymb]{revtex4}
\usepackage{float}
\usepackage{graphics,epsfig}
\usepackage{graphicx}
\usepackage{dcolumn}
\usepackage{bm}

\begin{document}
\baselineskip=0.45 cm

\title{Entanglement Enhanced Thermometry in the Detection of the Unruh Effect }

\author{Zehua Tian}
\email{zehuatian@126.com}
\affiliation{Institute of Theoretical Physics, University of Warsaw, Pasteura 5, 02-093 Warsaw, Poland}

\author{Jieci Wang}
\affiliation{Department of Physics, Key Laboratory of Low Dimensional Quantum Structures and Quantum Control \\
of Ministry of Education, and Synergetic Innovation Center for Quantum Effects and Applications, Hunan Normal University, 
Changsha, Hunan 410081, P. R. China}

\author{Jiliang Jing}
\email{jljing@hunn.edu.cn}
\affiliation{Department of Physics, Key Laboratory of Low Dimensional Quantum Structures and Quantum Control \\
of Ministry of Education, and Synergetic Innovation Center for Quantum Effects and Applications, Hunan Normal University, 
Changsha, Hunan 410081, P. R. China}

\author{Andrzej Dragan}
\email{dragan@fuw.edu.pl}
\affiliation{Institute of Theoretical Physics, University of Warsaw, Pasteura 5, 02-093 Warsaw, Poland}

\begin{abstract}

We show how the use of entanglement can enhance the precision of the detection of the Unruh effect with an accelerated probe. We use a two-level atom interacting relativistically with a quantum field as the probe, and treat it as an open quantum system to derive the master equation governing its evolution. By means of quantum state discrimination, we detect the accelerated motion of the atom by examining its time evolving state. It turns out that the optimal strategy for the detection of the Unruh effect, to which the accelerated atom is sensitive, involves letting the atom-thermometer equilibrate with the thermal bath. However, introducing initial entanglement between the detector and an external degree of freedom leads to an enhancement of the sensitivity of the detector. Also, the maximum precision is attained within finite time, before equilibration takes place.

\end{abstract}

\maketitle
\newpage
\section{Introduction}\label{section1}

Quantum field theory predicts that a relativistically accelerated observer views the Minkowski vacuum as a thermal bath with a temperature proportional to the proper acceleration of the observer. This phenomenon, known as the Unruh effect \cite{Unruh1}, is extremely weak, and even the acceleration as high as $a\sim10^{21}\text{ms}^{-2}$ should result in the temperature as low as $1\mathrm{K}$. Although several experimental proposals have been put forth \cite{Bell, Rogers, Unruh2, Chen, Matsas, Vanzella, Scully, Schutzhold, Martin, Rad, Felicetti,Retzker,Nori}, verification of the Unruh effect still remains an open question \cite{Crispino, John, Igor}.

The theory of the Unruh effect has attracted widespread interest in the physics community and it has been extensively examined from different perspectives. These include the approaches involving Bogolyubov transformation between the Minkowski and Rindler frames of reference \cite{Unruh1, Crispino, Dragan1, Dragan2, Doukas2, Ahmadi}, the application of the Unruh-DeWitt detector model which offers an operational definition of a particle as a trigger for the transitions between the atomic energy levels \cite{Unruh1, Birrell}, the manifestation of thermalization of open quantum systems involving decoherence and dissipation \cite{Benatti}, understanding as a quantum noise channel \cite{Omkar}, and the thermal correction to the energy shift and spontaneous excitation of atoms \cite{Dalibard,Audretsch, Yu}. In recent years, the Unruh effect has been extensively considered in the field of quantum information. It has been studied how the Unruh effect affects quantum entanglement between different observers \cite{Dragan1, Dragan2, Doukas2, Ahmadi,Landulfo1, Alsing2, Doukas, Bruachi1, Gerardo1, Wang1, Hwang, Bruschi2, Friis1}, as well as other types of non-classical correlations \cite{Doukas2,Datta, Wang2, Celeri, Tian1, Brown, Gerardo2, Qiang} and consequently quantum nonlocality \cite{Friis2, Smith, Dawkil, Tian2}, quantum teleportation \cite{Alsing1, Landulfo1}, entropic uncertainty relations \cite{Feng}, and parameter estimation \cite{Yao, Zahid, Hao}. The Unruh effect has also been used to analyze how to improve information processing protocols \cite{Friis3, Martin1} and quantum measurement techniques \cite{Mehdi}. 

In order to detect the Unruh effect several metrological tools have been employed. Different approaches include using Gaussian probes \cite{Aspachs}, applying the framework of open quantum systems to compute Fisher information and quantum Fisher information \cite{Tian3}, and using the technique of quantum channel discrimination \cite{Doukas1}. 

In this work, we study an overlooked possibility of increasing the detection precision for the Unruh effect and show how entangled states can be used to enhance sensibility of the accelerated detectors. We use quantum state discrimination technique \cite{Multi-Authors} to search for the optimal strategies that can facilitate the detection of the Unruh effect. We employ the model of a two-level atom coupled to the massless scalar fields in the Minkowski vacuum state as a probe and study its time evolving state to discriminate between two possible scenarios: one, in which the atom has been inertial, and the other, in which the atom has been uniformly accelerated. We discuss different strategies of detection and show how initial entanglement between the atom and an external degree of freedom can enhance the performance of the atom as a thermometer. 

Our paper is constructed as follows: in section \ref{section2} we introduce the model of our probe, we analyze the optimal detection of the Unruh effect in section \ref{section3}, and section \ref{section4} concludes the paper.


\section{Detection model} \label{section2}

Let us note that the detector model we will consider was firstly introduced in Ref. \cite{Benatti} by F. Benatti and R. Floreanin, where the Unruh effect has been reexamined from the perspective of the thermalization phenomenon and entanglement generation of accelerated atoms. Recently this model has been fruitfully applied to understand the Hawking and Gibbons-Hawking effects \cite{Yu1}. Here we will use this model to investigate the detection of the Unruh effect with the assistance of entanglement.

A combined system consists of a detector and an external fluctuating vacuum field with the total Hamiltonian of the general form:
\begin{eqnarray}\label{Hamiltonian}
H=H_D+H_\Phi+H_I,
\end{eqnarray}
where $H_D$ is the Hamiltonian of the detector, $H_\Phi$ is the free Hamiltonian of the scalar field, and $H_I$ represents their mutual interaction. 
The detector will be described with a two-level atom governed by the Hamiltonian $H_D=\frac{1}{2}\omega_0\sigma_z$, where $\omega_0$ is its energy-level spacing and $\sigma_z$ is the Pauli matrix. The interaction Hamiltonian takes the form:
\begin{equation}
H_I=\mu(\sigma_++\sigma_-)\Phi(x(\tau)),
\end{equation}
where $\mu$ is the coupling constant, $\sigma_+$ ($\sigma_-$) is the atomic raising (lowering) operator, and $\Phi(x)$ corresponds to the scalar field operator taken at $x(\tau)$ representing the classical trajectory of the atom.

We assume that the initial state of the atom, $\rho(0)$, is separable from the initial state of the field, which is assumed to be in the Minkowski vacuum $|0\rangle_\text{M}$. Therefore the initial state of the total system is given by $\rho_{tot}=\rho(0)\otimes|0\rangle_\text{M}\langle 0|$. In the reference frame of the atom, the equation of motion of the total system is given by:
\begin{eqnarray}\label{motion equation}
\frac{\text{d}\rho_{tot}(\tau)}{\text{d}\tau}=-i[H,\rho_{tot}(\tau)],
\end{eqnarray}
where $\tau$ is the proper time along the trajectory of the atom, and $\rho_{tot}(\tau)$ denotes the time-dependent density matrix of the atom and the field. Generally,
we will be interested in the evolution of the atom and this dynamics can be obtained by tracing out the field degrees of freedom, i.e., $\rho(\tau)=\text{Tr}_\Phi[\rho_{tot}(\tau)]$.
The reduced state density of the atom, in the limit of weak coupling, is found to obey the Kossakowski-Lindblad equation of the form \cite{Benatti,Gorini}:
\begin{eqnarray}\label{Lindblad equation}
\frac{\text{d}\rho(\tau)}{\text{d}\tau}&=&-i[H_{eff},\rho(\tau)]+\mathcal{L}[\rho(\tau)]
\end{eqnarray}
with the effective Hamiltonian $H_{eff}$ given below, and the Lindblad superoperator:
\begin{eqnarray}\label{Loperator}
\mathcal{L}[\rho(\tau)]=\frac{1}{2}\sum^3_{i,j=1}a_{ij}[2\sigma_j\rho\sigma_i-\sigma_i\sigma_j\rho-\rho\sigma_i\sigma_j].
\end{eqnarray}
Both the matrix $a_{ij}$ and the effective Hamiltonian $H_{eff}$ are associated with the field correlation functions, $G^+(x-x')=\mu^2\langle0|\Phi(x)\Phi(x')|0\rangle$.
By defining the Fourier and Hilbert transforms of the field correlation functions, namely:
\begin{eqnarray}\label{FHT}
\mathcal{G}(\lambda)&=&\int^{\infty}_{-\infty}d\tau e^{i\lambda\tau}G(x(\tau)),
\\
\mathcal{K}(\lambda)&=&\frac{P}{\pi i}\int^{\infty}_{-\infty}d\omega\frac{\mathcal{G}(\omega)}{\omega-\lambda},
\end{eqnarray}
we have computed that the coefficients of the Kossakowski matrix $a_{ij}$ can be expressed as:
\begin{eqnarray}\label{amatrix}
a_{ij}=A\delta_{ij}-iB\epsilon_{ijk}\delta_{k3}-A\delta_{i3}\delta_{j3},
\end{eqnarray}
with $A=\frac{1}{4}[\mathcal{G}(\omega_0)+\mathcal{G}(-\omega_0)]$, $B=\frac{1}{4}[\mathcal{G}(\omega_0)-\mathcal{G}(-\omega_0)]$.
The effective Hamiltonian $H_{eff}$, which contains the so called \emph{Lamb shift} \cite{Lamb}, then can be written as:
\begin{equation}\label{EH}
H_{eff}=\frac{1}{2}\Omega\sigma_z=\frac{1}{2}\left\{\omega_0+\frac{i}{2}[\mathcal{K}(-\omega_0)-\mathcal{K}(\omega_0)]\right\}\sigma_z,
\end{equation}
where we have defined a renormalized energy level spacing $\Omega$, which contains the original atomic energy level spacing $\omega_0$ and the energy shift $\frac{i}{2}[\mathcal{K}(-\omega_0)-\mathcal{K}(\omega_0)]$. Assuming the atom is initially prepared in the pure state $|\psi(0)\rangle=\sin\frac{\theta}{2}|0\rangle+\cos\frac{\theta}{2}|1\rangle$, by solving the master equation (\ref{Lindblad equation}) we can obtain the time-dependent reduced 
density matrix of the atom,
\begin{widetext}
\begin{eqnarray}\label{reduced-matrix}
\rho(\tau)=\left(
\begin{array}{cc}
e^{-4A\tau}\cos^2\frac{\theta}{2}+\frac{B-A}{2A}(e^{-4A\tau}-1) & \frac{1}{2}e^{-2A\tau-i\Omega\tau}\sin\theta \\ 
 \frac{1}{2}e^{-2A\tau+i\Omega\tau}\sin\theta  & 1-e^{-4A\tau}\cos^2\frac{\theta}{2}-\frac{B-A}{2A}(e^{-4A\tau}-1)
\end{array}
\right).
\end{eqnarray}
\end{widetext}

Let us now consider the evolution of the uniformly accelerated two-level atom, whose trajectory can be described as:
\begin{eqnarray}\label{trajectory}
t(\tau)=\frac{1}{a}\sinh(a\tau),~~x(\tau)=\frac{1}{a}\cosh(a\tau),~~y(\tau)=z(\tau)=0.
\end{eqnarray}
In order to obtain the parameters of the reduced density matrix of the accelerated atom, we need to evaluate the correlation function of the field, which can be found from the following two-point function of the massless scalar field:
\begin{eqnarray}\label{two-point-function}
G^+(x, x^\prime)=-\frac{\mu^2}{4\pi^2}\frac{1}{(t-t^\prime-i\epsilon)^2-|\mathbf{x}-\mathbf{x}^\prime|^2}.
\end{eqnarray}
Substituting the trajectory of the atom (\ref{trajectory}) into (\ref{two-point-function}), one can obtain the field correlation function in the frame of the accelerated atom \cite{Birrell}:
\begin{eqnarray}\label{wightman-function}
G^+(x, x^\prime)=-\frac{a^2\mu^2}{16\pi^2}\sinh^{-2}\bigg[\frac{a(\tau-\tau^\prime)}{2}-i\epsilon\bigg].
\end{eqnarray}
In this case, the Fourier transform of the field correlation function \eqref{FHT} is:
\begin{eqnarray}\label{FT}
\mathcal{G}(\lambda)=\frac{\mu^2\lambda}{4\pi}\bigg(1+\coth\frac{\pi\lambda}{a}\bigg).
\end{eqnarray}
Consequently, the coefficients of the Kossakowski matrix $a_{ij}$ and the effective level spacing $\Omega$ of the atom are given by:

\begin{eqnarray}\label{KM}
\nonumber
A&=&\frac{1}{4}\gamma_0\left(\frac{e^{2\pi\omega_0/a}+1}{e^{2\pi\omega_0/a}-1}\right),\nonumber \\
B &=&\frac{1}{4}\gamma_0,
\nonumber \\
\Omega&=&\omega_0+\frac{\gamma_0}{2\pi\omega_0}\int^\infty_{-\infty}\text{d}\omega\,
P\left(\frac{\omega}{\omega+\omega_0}-\frac{\omega}{\omega-\omega_0}\right)
\nonumber \\
&&\times\left(1+\frac{2}{e^{2\pi\omega/a}-1}\right),
\end{eqnarray}
where $\gamma_0=\mu^2\omega_0/2\pi$ is the spontaneous emission rate of the inertial atom in Minkowski spacetime. Using the above results and assuming that the time of evolution is sufficiently long, i.e. $\tau\gg1/A$, one can write the steady state of the accelerated atom as:
\begin{eqnarray}\label{thermal-state}
\rho(\infty)=\frac{e^{\beta\,H_D}}{Tr[e^{\beta\,H_D}]},
\end{eqnarray}
which is a thermal state with the temperature $T=1/\beta=a/2\pi$. This result shows that the accelerated atom has been equilibrating with the thermal bath
characterized by the mean number of Unruh particles equal to $N_U=1/(e^{\frac{\omega_0}{T}}-1)$. For further convenience, let us parameterize 
the number of Unruh particles by $n=1+2N_U$. Then, in the interaction representation we can rewrite the time-dependent state of the accelerated atom as:
\begin{eqnarray}\label{bloch-representation}
\nonumber
\rho(\tau)&=&\frac{1}{2}\bigg(1+\sum^3_{i=1}r_i(\tau)\sigma_i\bigg),
\\ \nonumber
r_1(\tau)&=&r_1(0)e^{-\frac{1}{2}\gamma_0\,n\tau},~~~~~~r_2(\tau)=r_2(0)e^{-\frac{1}{2}\gamma_0\,n\tau},
\\
r_3(\tau)&=&r_3(0)e^{-\gamma_0\,n\tau}-\frac{1}{n}(1-e^{-\gamma_0\,n\tau}),
\end{eqnarray}
where $r_1(0)=\sin\theta$, $r_2(0)=0$ and $r_3(0)=\cos\theta$. Let us note that if $n=1$, then Eq. (\ref{bloch-representation}) represents the the evolving quantum state of the inertial atom in the Minkowski vacuum.


\section{Optimally detecting the Unruh effect} \label{section3}

We will now use the detector model described in the previous section to act as a thermometer detecting the Unruh effect. In order to find out how to optimally detect this effect, we choose to measure the atom in the basis that has the lowest probability of misidentifying the two possible states: one characterizing a static, and the other - uniformly accelerated trajectory of the atom. It is clear that this question has been reduced to the classical problem of quantum state discrimination \cite{Multi-Authors}. In this scheme, the minimum probability of making an error in the identification of the states $\rho_1$ or $\rho_2$ is given by: 
\begin{eqnarray}\label{error-p1}
P_E=\frac{1}{2}(1-\frac{1}{2}||\rho_1-\rho_2||),
\end{eqnarray}
where $||\Delta||=Tr\sqrt{\Delta^\dagger\Delta}$ denotes the trace norm of $\Delta$. For Hermitian operators $\Delta$ their norms also equal the sum of the absolute values of their eigenvalues. In our problem, $\rho_1$ is the evolution state of the inertial atom, i.e., the case of $n=1$ in Eq.~\eqref{bloch-representation}, and $\rho_2$ corresponds to the quantum state of the accelerated atom, i.e., the case of $n>1$ in Eq.~\eqref{bloch-representation}. From this, we obtain:
\begin{eqnarray}\label{error-p2}
P_E=\frac{1}{2}\bigg(1-\frac{1}{2}\Delta(\vec{r}_1(\tau)-\vec{r}_2(\tau))\bigg)
\end{eqnarray}
with
\begin{eqnarray}\label{delta}
\Delta^2(\vec{r}_1(\tau)-\vec{r}_2(\tau))=\sin^2\theta\Lambda^2_1
+\bigg[\cos\theta\Lambda_2-\Lambda_3\bigg]^2,
\end{eqnarray}
where $\Lambda_1=e^{-\frac{1}{2}\gamma_0\tau}-e^{-\frac{1}{2}\gamma_0n\tau}$, $\Lambda_2=e^{-\gamma_0\tau}-e^{-\gamma_0n\tau}$, and
 $\Lambda_3=(1-e^{-\gamma_0\tau})-\frac{1}{n}(1-e^{-\gamma_0n\tau})$. For sufficiently long interaction times, the atom reaches equilibrium with the state of the external field, then the distinguishability between the two equilibrium states in Eq. (\ref{delta}) is determined by $\Delta_\infty=1-n^{-1}$. Moreover, in the long limit of interaction times the level of distinguishability is independent of the initial state of such a thermometer.

\begin{figure}[ht]
\centering
\includegraphics[width=0.46\textwidth]{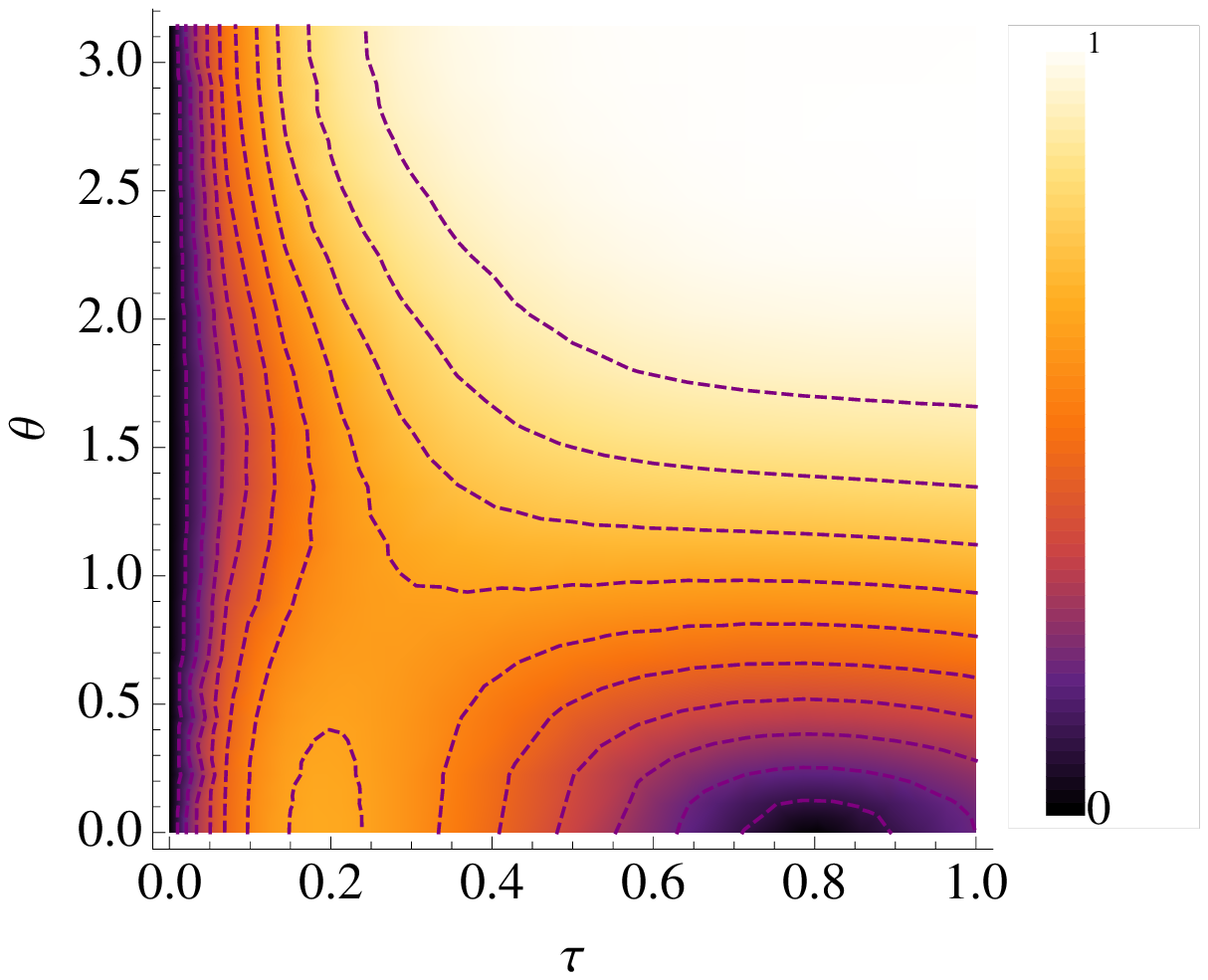}
\includegraphics[width=0.42\textwidth]{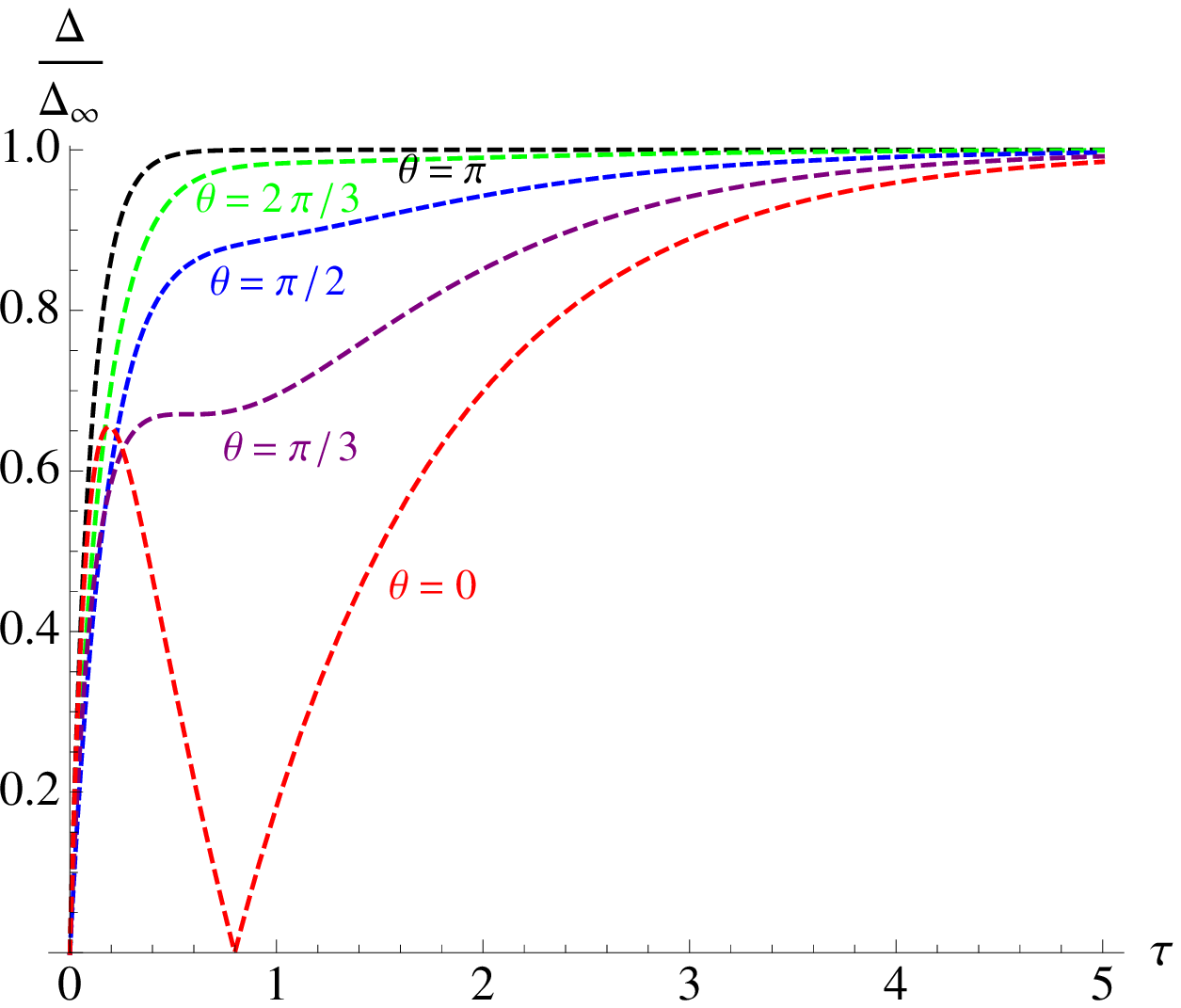}
\caption{(Color online) The Euclidean distance $\Delta(\vec{r}_1(\tau)-\vec{r}_2(\tau))$, normalized to $\Delta_\infty$, plotted as a function of time $\tau$ (in units of $\gamma_0$) and the parameter $\theta$ characterizing the initial state of the atom. $n=10$ is assumed.}\label{fig1}
\end{figure}

Let us now study the distinguishability of the state for finite interaction times. To better quantify the performance of the atom thermometer, let us normalize the Euclidean distance $\Delta(\vec{r}_1(\tau)-\vec{r}_2(\tau))$ to $\Delta_\infty$. In Fig. \ref{fig1}, we plot $\Delta(\vec{r}_1(\tau)-\vec{r}_2(\tau))/\Delta_\infty$ as a function of the interaction time $\tau$ and the parameter $\theta$ characterizing the initial state of the atom thermometer. We can see from Fig.~\ref{fig1} that the two possible trajectories are always less distinguishable as compared to the equilibrium case. The largest distance between the two alternative final states is obtained in the limit of long interaction times, and it is equal to $1-n^{-1}$. Interestingly the largest distance in this case is independent of initial states of the atom thermometer. It means that the optimal condition to detect the Unruh effect is letting the atom evolve for long enough times regardless of its initial state. 
Thus, we arrive at the conclusion that quantum coherence \cite{Baumgratz} plays a trivial role in the optimal detection of the Unruh effect. Let us note that our results obtained here are consistent with that of Ref.~\cite{Tian3} where the technique of quantum metrology has been used. 

We also find that if the probe is prepared in the excited state, i.e., $\theta=0$, then there is a finite time, $\tau\approx0.80$ (namely, when $\Lambda_2=\Lambda_3$), at which the trace distance goes to zero. This is because at this time, both the accelerated atom and the inertial atom are approximatively driven to the same thermal state with the parameterized temperature $n=10$. However, after that time, the state of the accelerated atom remains almost unchanged, while the state of the inertial atom will be eventually driven to the ground state. Thus, we can see that the Euclidean distance between these two cases will increase again and finally reach $1-n^{-1}$. We can also see from Fig.~\ref{fig1} that the increase of $\theta$ leads to a quicker equilibration of the system.

We will now turn our attention to a more general scenario, in which the considered atom is initially entangled with another atom. In particular we assume that the additional atom does not interact with the external field. Let us consider the initial state of the two atoms to be:
\begin{equation}
\rho(0)=\frac{1}{4}\bigg(1+\sum^3_{i=1}c_i(0)\sigma_i\otimes\sigma_i\bigg).
\end{equation} 
Note that this state is an $X-$type state and the parameter $c_i$ satisfies $0\le|c_i|\le1$; the two particular cases, $|c_i|=1$ and $|c_i|=c$ correspond to one of the Bell states and the Werner state, respectively. Then the evolution of the system leads to the following final state of the atoms:
\begin{eqnarray}\label{two-time-dependent-state}
\nonumber
\rho(n, \tau)&=&\frac{1}{4}\bigg(1+\sum^3_{i=1}c_i(n, \tau)\sigma_i\otimes\sigma_i+c_{30}(n, \tau)\sigma_3\otimes\sigma_0\bigg),
\\ \nonumber
c_1(n, \tau)&=&c_1(0)e^{-\frac{1}{2}\gamma_0\,n\tau},~~~~~~c_2(n, \tau)=c_2(0)e^{-\frac{1}{2}\gamma_0\,n\tau},
\\
c_3(n, \tau)&=&c_3(0)e^{-\gamma_0\,n\tau},~~~~c_{30}(n, \tau)=-\frac{1}{n}(1-e^{-\gamma_0\,n\tau}).
\end{eqnarray}
With this in hand, we can compute the trace distance between the state $\rho(1, \tau)$ corresponding to the resting detector case and the state $\rho(n, \tau)$ corresponding to the accelerated detector case, which is given by:
\begin{eqnarray}\label{trace-distance}
\nonumber
\Delta&=&\frac{1}{4}\bigg[\bigg|\sqrt{c^2_+(0)\Lambda^2_1+\Lambda^2_3}+c_3(0)\Lambda_2\bigg|+\bigg|\sqrt{c^2_+(0)\Lambda^2_1+\Lambda^2_3}
\\  \nonumber
&&-c_3(0)\Lambda_2\bigg|+\bigg|c_3(0)\Lambda_2+\sqrt{c^2_-(0)\Lambda^2_1+\Lambda^2_3}\bigg|+\bigg|c_3(0)\Lambda_2
\\ 
&&-\sqrt{c^2_-(0)\Lambda^2_1+\Lambda^2_3}\bigg|\bigg],
\end{eqnarray}
where $c_\pm(0)=c_1(0)\pm\,c_2(0)$. Let us note that in the limit of infinite evolution time, the trace distance also approaches to $1-n^{-1}$ regardless of the initial state of the two atoms.

\begin{figure}[ht]
\centering
\includegraphics[width=0.45\textwidth]{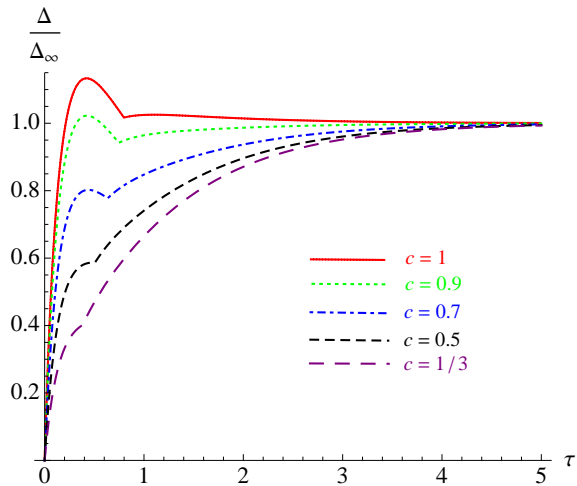}
\caption{(Color online) Normalized trace distance for bipartite states as a function of time $\tau$ (in units of $\gamma_0$). $n=10$ is taken.}\label{fig2}
\end{figure}

Consider the Werner initial state, for which $c_1(0)=-c_2(0)=c_3(0)=c$ with $0\leq\,c\leq1$. Such an input state is only entangled when $1/3<c\le1$. The resulting trace distance given by Eq.~\eqref{trace-distance} is plotted in Fig.~\ref{fig2}. We find that the trace distance between the accelerated and non-accelerated scenario increases as a function of the initial entanglement (quantified by $c$). Besides, for some entangled states there is a global maximum in the distinguishability before they achieve the equilibrium with the environment, and interestingly this value can even exceed the equilibrium value, $1-n^{-1}$. This interesting result proves that the performance of the thermometer with the auxiliary component can be enhanced with the aid of the initial entanglement. It also follows that the optimal bipartite state for the detection of Unruh effect is the maximally entangled state ($c=1$).

Let us now elucidate some of the phenomena observed in Fig.~\ref{fig2}. There is a sudden change point for the trace distance, which occurs when $c\Lambda_2=\Lambda_3$. Thus, the bigger $c$ is, the later the sudden change occurs. If $c>\inf_\tau\bigg\{\frac{\sqrt{16\Lambda^2_1(1-n^{-1})^2+\Lambda^2_2\Lambda^2_3-4\Lambda^2_1\Lambda^2_3}-2\Lambda_2(1-n^{-1})}{4\Lambda^2_1-\Lambda^2_2}\bigg\}$ (it is equal to $0.88$ for $n=10$ in Fig.~\ref{fig2}), the trace distance can become larger than the equilibrium one, $1-n^{-1}$. In this case there is also the maximum trace distance corresponding to the the optimal detection of the acceleration of the atom. In the case of $c=1$ ($c=0.9$), the maximum trace distance is $1.13$ ($1.02$) when $\tau=0.42$ ($\tau=0.43$).

\begin{figure}[ht]
\centering
\includegraphics[width=0.45\textwidth]{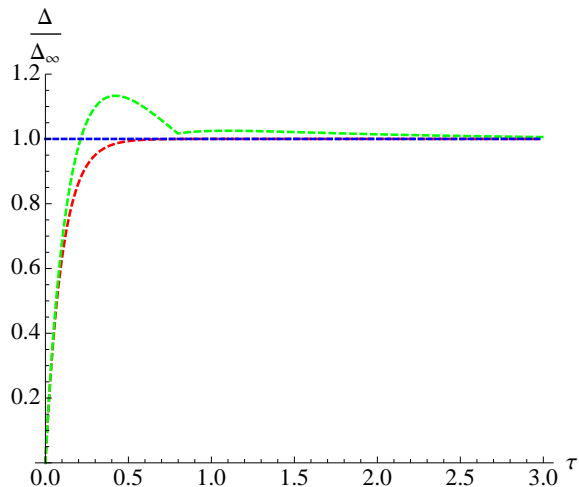}
\caption{(Color online) Normalized trace distances as a function of time $\tau$ (in units of $\gamma_0$). The green curve denotes that for the initial maximally entangled state case, the red curve denotes that for the single atom initially prepared in the ground state, and the blue curve denotes the equilibrium case. $n=10$ is assumed here.}\label{fig3}
\end{figure}

To find out which thermometer, the one with or without the initial entanglement, performs better in the detection of the Unruh effect, we compare the trace distance of bipartite system case with that of a single atom case in Fig.~\ref{fig3}. It is clear that the use of the initial entanglement provides a significant advantage. This interesting result provides a hint that the use of entanglement can be useful in the detection of the absolute acceleration, and as a consequence in experimental verification of the Unruh effect.

\section{Conclusions}\label{section4}

We have used a simple model of a two-level atom interacting relativistically with the fluctuating massless scalar quantum field in the Minkowski vacuum to probe the Unruh effect. The purpose of the study was to determine the optimum methods of the detection, in particular to show a relevant role of the initial entanglement, which has been overlooked in the previous studies. Within the framework of open quantum systems, we have obtained the analytical dynamical evolution of the thermometer. We have then compared the evolution of the accelerated atom with that of the inertial one using the quantum state discrimination techniques. We have found that regardless of the initial state of the atom, the largest trace distance between this two cases is $1-n^{-1}$, which is achieved in the limit of infinitely long interaction times. This results corresponds to the situation when the probe equilibrates with the surrounding thermal bath of the field. However, we have shown that this result can be improved by considering a more general situation in which the atom thermometer is initially entangled with another two-level system that does not interact with the considered field. Our results shows that the elusive Unruh effect, which has been too difficult to be experimentally verified so far, perhaps can be better explored with the aid of entangled systems in accelerated motion.

Let us note that our method here, based the equivalence principle, can also be applied to the detection of Hawking effect \cite{Hawking} and Gibbons-Hawking effect \cite{Gibbons}, and so on. Besides, with the development of simulation experiments of relativistic quantum systems \cite{Marco, Felicetti}, our results can be relevant to upcoming experimental tests of related phenomena in analogue systems.

\begin{acknowledgments}
Z. Tian and A. Dragan thank for the financial support to the National Science Center, Sonata BIS Grant No. DEC-2012/07/E/ST2/01402. J. Jing and J. Wang are supported by the National Natural Science Foundation of China under Grant Nos. 11475061, and 11305058.
\end{acknowledgments}



\begin{thebibliography}{99}
\bibitem{Unruh1}
W. G. Unruh, Phys. Rev. D {\bf 14}, 870 (1976).

\bibitem{Bell}
J. S. Bell and J. M. Leinaas, Nucl. Phys. B {\bf 212} (1983) 131; J. S. Bell and J. M. Leinaas, Nucl. Phys. B {\bf 284} (1987) 488.

\bibitem{Rogers}
J. Rogers, Phys. Rev. Lett. {\bf 61}, 2113 (1988).

\bibitem{Unruh2}
W. G. Unruh, Phys. Rep. {\bf 307} (1998) 163-171.

\bibitem{Chen}
P. Chen and T. Tajima, Phys. Rev. Lett. 83, {\bf 256} (1999).

\bibitem{Matsas}
G.E.A. Matsas and D.A.T. Vanzella, Phys. Rev. D {\bf 59}, 094004 (1999).

\bibitem{Vanzella}
D. A. T. Vanzella and G. E. A. Matsas, Phys. Rev. Lett. {\bf 87}, 151301 (2001).

\bibitem{Scully}
M. O. Scully ,V.V. Kocharovsky,A. Belyanin, E. Fry, F. Capasso, Phys. Rev. Lett. {\bf 91}, 243004 (2003).

\bibitem{Schutzhold}
R. Sch\"{u}tzhold, G. Schaller, and D. Habs, Phys. Rev. Lett. {\bf 97}, 121302 (2006).

\bibitem{Martin}
E. Mart\'{\i}nÐMart\'{\i}nez, I. Fuentes, and R. B. Mann, Phys. Rev. Lett. {\bf 107}, 131301 (2011).

\bibitem{Rad}
N. Rad and D. Singleton, Eur. Phys. J. D (2012) {\bf 66}: 258.

\bibitem{Felicetti}
S. Felicetti, C. Sab\'{\i}n, I. Fuentes, L. Lamata, G. Romero, and E. Solano, Phys. Rev. B {\bf 92}, 064501 (2015).


\bibitem{Retzker}
A. Retzker, J.I.Cirac, M.B. Plenio, B. Reznik, Phys. Rev. Lett. {\bf 101}, 110402 (2008).

\bibitem{Nori}
P. D. Nation, J. R. Johansson, M. P. Blencowe, F. Nori Rev. Mod., Phys. {\bf 84}, 1 (2012).

\bibitem{Crispino}
L.C.B. Crispino, A.Higuchi, G. E.A. Matsas, Rev. Mod. Phys. {\bf 80}, 787 (2008).


\bibitem{John}
J. Earman, Studies in History and Philosophy of Modern Physics {\bf 42} (2011) 81-97.

\bibitem{Igor}
I. Pe\~{n}a, and D. Sudarsky, Found Phys (2014) {\bf 44}: 689-708.


\bibitem{Dragan1}
A. Dragan, J. Doukas, E. Martin-Martinez, and DE Bruschi, Class. Quantum Grav. {\bf 30}, 235006 (2013).

\bibitem{Dragan2}
A. Dragan, J. Doukas, and E. Martin-Martinez, Phys. Rev. A {\bf 87} 052326 (2013).

\bibitem{Doukas2}
J. Doukas, E. G. Brown, A. Dragan, and R. B. Mann, Phys. Rev. A {\bf 87}, 012306 (2013).

\bibitem{Ahmadi}
M. Ahmadi, K. Lorek, A. Checinska, A. R. H. Smith, R. B. Mann, and A. Dragan, arXiv:1602.02349 [quant-ph] (2016). 

\bibitem{Birrell}
N. D. Birrell, P. C. W. Davies, \emph{Quantum Fields in Curved Space} (Cambridge University Press, Cambridge, 1982).

\bibitem{Benatti}
F. Benatti and R. Floreanini, Phys. Rev. A {\bf 70}, 012112 (2004).

\bibitem{Omkar}
S. Omkar, S. Banerjee, R. Srikanth, and A. K. Alok, Quantum Inf. and Comp. {\bf 16} (2016) 0757; S. Banerjee, A. K. Alok, S. Omkar, and R. Srikanth, arXiv:1603.05450 [quant-ph].

\bibitem{Dalibard}
J. Dalibard, J. Dupont-Roc, and C. Cohen-Tannoudji, J. Phys. (France) {\bf 43}, 1617 (1982);
J. Dalibard, J. Dupont-Roc, and C. Cohen-Tannoudji, J. Phys. (France) {\bf 45}, 637 (1984).

\bibitem{Audretsch}
J. Audretsch, and R. M\"{u}ller, Phys. Rev. A {\bf 52}, 629 (1995);
Phys. Rev. A {\bf 50}, 1755 (1994).

\bibitem{Yu}
Zh. Zhu, H. Yu, and Sh. Lu, Phys. Rev. D {\bf 73}, 107501 (2006).

\bibitem{Landulfo1}
A. G. S. Landulfo, and G. E. A. Matsas, Phys. Rev. A {\bf 80}, 032315 (2009).

\bibitem{Alsing2}
P. M. Alsing, I. Fuentes-Schuller, R. B. mann, and T. E. Tessier, Phys. Rev. A {\bf 74}, 032326 (2006).

\bibitem{Doukas}
J. Doukas, and L. C. L. Hollenberg, Phys. Rev. A {\bf 79}, 052109 (2009).

\bibitem{Bruachi1}
D. E. Bruschi. J. Louko, E. Mar\'{\i}n-Mart\'{\i}nez, A. Dragan, and I. Fuentes, Phys. Rev. A {\bf 82}, 042332 (2010).

\bibitem{Gerardo1}
G. Adesso, I. Fuentes-Schuller, and M. Ericsson, Phys. Rev. A {\bf 76}, 062112 (2007).

\bibitem{Wang1}
J. Wang, and J. Jing, Phys. Rev. A {\bf 83}, 022314 (2011).

\bibitem{Hwang}
M. R. Hwang, D. Park, and E. Jung, Phys. Rev. A {\bf 83}, 012111 (2011).

\bibitem{Bruschi2}
D. E. Bruschi, I. Fuentes, and J. Louko, Phys. Rev. D {\bf 85}, 061701(R) (2012).

\bibitem{Friis1}
N. Friis, A. R. Lee, D. E. Bruschi, and J. Louko, Phys. Rev. D {\bf 85}, 025012 (2012).

\bibitem{Datta}
A. Datta, Phys. Rev. A {\bf 80}, 052304 (2009).

\bibitem{Wang2}
 J. Wang, J. Deng, and J. Jing, Phys. Rev. A {\bf 81}, 052120 (2010).
 
\bibitem{Celeri}
L. C. C\'{e}leri, A. G. S. Landulfo, R. M. Serra, and G. E. A. Matsas, Phys. Rev. A {\bf 81}, 062130 (2010).
 
 \bibitem{Tian1}
 Z. Tian, and J. Jing, Physics Letters B {\bf 707} (2012) 264-271.
 
 \bibitem{Brown}
 E. G. Brown, K. Cormier, E. Mart\'{\i}-Mart\'{\i}nez, and R. B. Mann, Phys. Rev. A {\bf 86}, 032108 (2012).
 
 \bibitem{Gerardo2}
 G. Adesso, S. Ragy, and D. Girolami, Class. Quantum Grav. {\bf 29} (2012) 224002 (19pp).
 
 \bibitem{Qiang}
 W. Qiang, and L. Zhang, Physics Letters B {\bf 742} (2015) 383-389.

\bibitem{Friis2}
N. Friis, P. K\"{o}hler, E. Mart\'{\i}n-Mart\'{\i}nez, anf R. A. Bertlmann, Phys. Rev. A {\bf 84}, 062111 (2011).

\bibitem{Smith}
 A. Smith, and R. B. Mann, Phys. Rev. A {\bf 86}, 012306 (2012).
 
\bibitem{Dawkil}
D.Park, J. Phys. A: Math. Theor. {\bf 45} (2012) 415308 (10pp).
 
\bibitem{Tian2} 
Z. Tian, and J. Jing, Annals of Physics {\bf 333} (2013) 76-89; Z. Tian, J. Wang, and J. Jing, Annals of Physics {\bf 332} (2013) 98-109.

\bibitem{Alsing1}
P. M. Asling, and G. J. Milburn, Phys.Rev.Lett. {\bf 91}, 180404 (2003) ;
P. M. Asling, D. McMahon, and G. J. Milburn, J. Opt. B: Quantum Semiclass. Opt. {\bf 6} (2004) S834-S843.

\bibitem{Feng}
J. Feng, Y. Zhang, M. D. Gould, and H. Fan, Physics Letters B {\bf 726} (2013) 527-532.

\bibitem{Yao}
Y. Yao, X. Xiao, L. Ge, X. Wang, and Ch. Sun, Phys. Rev. A {\bf 89}, 042336 (2014).

\bibitem{Zahid}
Z. H. Shamsi, D. G. Kim, and Y. H. kwon, arXiv:1409.6847 [quant-ph].

\bibitem{Hao}
X. Hao, and Y. Wu,  arXiv:1510.06515 [quant-ph].

\bibitem{Friis3}
N. Friis, M. Huber, I. Fuentes, and D. E. Bruschi, Phys. Rev. D {\bf 86}, 105003 (2012); D. E. Bruschi, A. Dragan, A. R. Lee, I. Fuentes, and J. Louko, Phys. Rev. Lett. {\bf 111}, 090504 (2013); D. E. Bruschi, J. Louko, D. Faccio, and I. Fuentes, New Journal of Physics {\bf 15} (2013) 073052 (12pp).

\bibitem{Martin1}
E. Mart\'{\i}n-Mart\'{\i}nez, D. Aasen, and A. Kempf, Phys. Rev. Lett. {\bf 110}, 160501 (2013); E. Mart\'{\i}n-Mart\'{\i}nez, and C. Sutherland, Physics Letters B {\bf 739} (2014) 74-82.

\bibitem{Mehdi}
M. Ahmadi, D. E. Bruachi, C. Sab\'{\i}n, G. Adesso, and I. Fuentes, Sci. Rep. {\bf 4}, 4996 (2014).

\bibitem{Aspachs}
M. Aspachs, G. Adesso, and I. Fuentes, Phys. Rev. Lett. {\bf 105}, 151301 (2010).

\bibitem{Tian3}
Z. Tian, J. Wang, H. Fan, and J. Jing, Sci. Rep. {\bf 5} 7946 (2015).

\bibitem{Doukas1}
J. Doukas, G. Adesso, S. Pirandola, and A. Dragan, Class. Quantum Grav. {\bf 32}, 035013 (2015).


\bibitem{Multi-Authors}
C. W. Helstrom, \emph{Quantum Detection and Estimation Theory} (Academic Press, New York, 1976);
 A. Chefles, Contemp. Phys. {\bf 41}, 401 (2000); V. Kargin, Ann. Stat. {\bf 33}, 959 (2005); M. Tsang, Phys. Rev. Lett. {\bf 108}, 170502 (2012).
 
 \bibitem{Yu1}
H. Yu and J. Zhang, Phys. Rev. D {\bf 77}, 024031 (2007); H. Yu, Phys. Rev. Lett. {\bf 106}, 061101 (2011).

\bibitem{Gorini}
V. Gorini, A. Kossakowski, and E. C. G. Surdarshan, J. Math. Phys. {\bf 17}, 821 (1976);
G. Lindblad, Commun. Math. Phys. {\bf 48}, 119 (1976).

\bibitem{Lamb}
W. E. Lamb, R. C. Retherford, Phys. Rev {\bf 72}, 241 (1947).

\bibitem{Baumgratz}
T. Baumgratz, M. Cramer, M. B. Plenio, Phys. Rev. Lett. {\bf 113}, 140401 (2014).

\bibitem{Hawking}
S. W. Hawking, Nature {\bf 248}, 30-31 (1974); Commun. Math. Phys. {\bf 43}, 199-220 (1975).

\bibitem{Gibbons}
G. W. Gibbons, and S. W. Hawking, Phys. Rev. D {\bf 15}, 2738 (1977).

\bibitem{Marco}
M. del Rey, D. Porras, and E. Mart\'{\i}n-Mart\'{\i}nez, Phys. Rev. A {\bf 85}, 022511 (2012).


\end{thebibliography}
\end{document}